\documentclass{optica-article}

\journal{opticajournal} 

\articletype{Research Article}
\usepackage{upgreek}
\usepackage{float}
\usepackage{graphicx}
\usepackage{amsmath}
\usepackage{ulem}
\usepackage{subfigure}
\usepackage{graphics}
\usepackage{upgreek}
\usepackage{caption}
\usepackage{cuted}
\usepackage[section]{placeins}
\usepackage{xcolor}
\usepackage{xpatch}
\usepackage{stfloats}
\begin{document}

\title{Compact quantum random number generator based on a laser diode and silicon photonics integrated hybrid chip}

\author{Xuyang Wang\authormark{1,2,3*}, Tao Zheng\authormark{1}, Yanxiang Jia\authormark{1}, Qianru Zhao\authormark{1}, Yu Zhang\authormark{1}, Yuqi Shi\authormark{1}, Ning Wang\authormark{1,2}, Zhenguo Lu\authormark{1,2}, Jun Zou\authormark{4}, and Yongmin Li\authormark{1,2,3*}}

\address{\authormark{1}State Key Laboratory of Quantum Optics and Quantum Optics Devices, Institute
of Opto-Electronics, Shanxi University, Taiyuan 030006, People’s Republic of China\\
\authormark{2}Collaborative Innovation Center of Extreme Optics, Shanxi University, Taiyuan
030006, People’s Republic of China\\
\authormark{3}Hefei National Laboratory, Hefei 230088, People’s Republic of China\\
\authormark{4}ZJU-Hangzhou Global Scientific and Technological Innovation Center, Zhejiang University, Hangzhou 311215, People’s Republic of China}

\email{\authormark
{*}wangxuyang@sxu.edu.cn

{*}yongmin@sxu.edu.cn} 


\begin{abstract*}In this study, a compact and low-power-consumption quantum random number generator (QRNG) based on a laser diode and silicon photonics integrated hybrid chip is proposed and verified experimentally. The hybrid chip’s size is 8.8×2.6×1 $\text{mm}^3$, and the power of entropy source is 80 mW. A common mode rejection ratio greater than 40 dB was achieved using an optimized 1×2 multimode interferometer structure. A method for optimizing the quantum-to-classical noise ratio is presented. A quantum-to-classical noise ratio of approximately 9 dB was achieved when the photoelectron current is 1 $\upmu$A using a balance homodyne detector with a high dark current GeSi photodiode. The proposed QRNG has the potential for use in scenarios of moderate MHz random number generation speed, with low power, small volume, and low cost prioritized.  

\end{abstract*}

\section{ INTRODUCTION}
Random numbers are key resources in the information age. They play an essential role in various applications, such as simulation, cryptography, and fundamental physical experiments\cite{RevModPhys.89.015004,2016Quantum,2006Random,Lim2015Quantum,liu2023experimental,2023High}. A quantum random number generator (QRNG) can produce true random numbers with characteristics of unpredictability, irreproducibility, and unbiasedness, guaranteed by the basic principles of quantum physics. Over the last two decades, various QRNG schemes have been proposed, such as QRNGs based on branching path\cite{2000A}, time of arrival\cite{2007Quantum,2009Photon,wahl2011ultrafast}, photon counting\cite{furst2010high,ren2011quantum}, attenuated pulse\cite{wei2009bias}, vacuum fluctuations\cite{gabriel2010generator,shen2010practical,symul2011real}, phase noise\cite{guo2010truly,qi2010high,jofre2011true}, amplified spontaneous emission\cite{williams2010fast}, Raman scattering\cite{bustard2011quantum}, and optical parametric oscillators\cite{marandi2011twin}. In particular, vacuum fluctuation-based QRNGs have several advantages\cite{bruynsteen2023100}. First, the source of entropy, i.e., vacuum noise, is readily available; thus, no bulky external components are required. Second, the inherent canceling of excess noise present in the local oscillator using balanced detection relaxes requirements on the laser and increases the system’s resilience against external perturbation\cite{gabriel2010generator}. Third, all optics devices can be integrated into chips. Because of these advantages, several silicon photonics chip-based QRNGs have been implemented\cite{bruynsteen2023100,raffaelli2018homodyne,tasker2021silicon,bruynsteen2021integrated,bai202118}, and a QRNG with 100-GHz-bps generation speed can be achieved\cite{bruynsteen2023100}. 

For photonics chip-based QRNGs, although there are several materials for photonics integrated circuits, such as silicon, doped silicon dioxide ($\text{SiO}_2$), indium phosphide (InP)\cite{abellan2016quantum}, gallium arsenide (GaAs), silicon nitride ($\text{Si}_3\text{Ni}_4$), and lithium niobate ($\text{LiNbO}_3$), silicon photonics integrated circuit technology is dominant. Silicon photonics, which uses silicon-on-insulator (SOI) wafers as semiconductor substrate materials, is compatible with complementary metal oxide semiconductor (CMOS) fabrication. This means that most standard CMOS manufacturing processes can be applied, and silicon photonics technology can monolithically integrate silicon electronics and photonics into the same platform\cite{siew2021review}. To achieve a small-volume, low-cost, high-performance QRNG, many researchers have employed silicon photonics integrated technology. Because there are no light sources in silicon due to its indirect bandgap, laser beams are guided into silicon photonics chips by grating or edge coupling in integrated QRNGs\cite{bruynsteen2023100,raffaelli2018homodyne,tasker2021silicon,bruynsteen2021integrated,bai202118}. To further improve the feasibility and practicability of such QRNGs, the coupling fiber array can be packaged with chips using a copper block\cite{bai202118}. In these studies, a bulky external laser and polarization controller were also used. Nevertheless, research is required to realize a more compact QRNG based on a silicon photonic chip, such as using hybrid or heterogeneous technology to integrate the light source on chips\cite{kaur2021hybrid} and integrating optical and electrical parts on one chip. 

High generation speed is a critical characteristic of QRNGs. However, not all systems require large random numbers. For example, in the aviation field, a 1-MHz/s speed is sufficient for use in airborne radar jump frequency and waveform modulation, with low power, small volume, and low cost prioritized. Photonics integrated circuit-based QRNGs seem adequate to meet these requirements. 

In this study, to realize a compact, low-power QRNG, a hybrid integrated chip comprising a III–V InP laser diode (LD) chip and a silicon photonics chip is proposed and verified experimentally. The two chips are packaged together via edge coupling. The LD output beam is directly guided into the silicon photonics chip to minimize the QRNG’s volume. An advantage of the proposed QRNG technique is that it generates a megabit quantum random number per second with little power and a small volume. No high-power LD driver, temperature controller, or balancing structures are required. A method for optimizing the quantum-to-classical noise ratio (QCNR), which transforms various noises into current noises, is presented. The theoretical analysis results are consistent with the experimental results. A voltage amplifier is used to flexibly tune the standard deviation (SD) of the output voltage noise according to the analog-to-digital converter (ADC) scales of the QRNG. A postprocessing module based on a field-programmable gate array (FPGA) is used to generate quantum random numbers in real time.

The remainder of this article is organized as follows. In Section 2, the structure of the proposed QRNG with a hybrid chip is introduced; the packing process is described in detail. In Section 3, we analyze the various noises in the proposed QRNG; a comprehensive comparison between the noises of balance homodyne detector (BHD) with commercial InGaAs photodiode and BHD with GeSi photodiode is presented, and the common mode rejection ratios (CMRR) are measured. In Section 4, the generation of quantum random numbers is introduced in detail. A method for flexibly tuning the SD of the output voltage according to the ADC scale is presented. Finally, Section 5 presents the conclusions and outlook.

\section{ QRNG STRUCTURE}
The proposed compact QRNG comprises three major parts (Fig. 1(a)): a hybrid chip, an analog circuit, and a digital circuit. The hybrid chip mainly comprises a 1550-nm InP edge-emitting LD chip, a silicon photonics chip, and two mounts. The LD is a mature commercial product, and the silicon photonics chip is fabricated using industry-standard active flow SOI technology CSiP180A1 of CUMEC; these are the QRNG’s optical parts. The two mounts are customized using aluminum nitride ceramics with good heat conduction and are used to pack the hybrid chip. The analog circuit mainly comprises a low-noise transimpedance amplifier (TIA) ADA4817, a low-noise voltage amplifier OPA847, a constant current (CC) source, and a high-pass filter (HPF). The  two parts above are noted as entropy source. The digital circuit mainly comprises an ADC and randomness extraction module based on the FPGA. 

\begin{figure}[ht]
\centering\includegraphics[width=12cm]{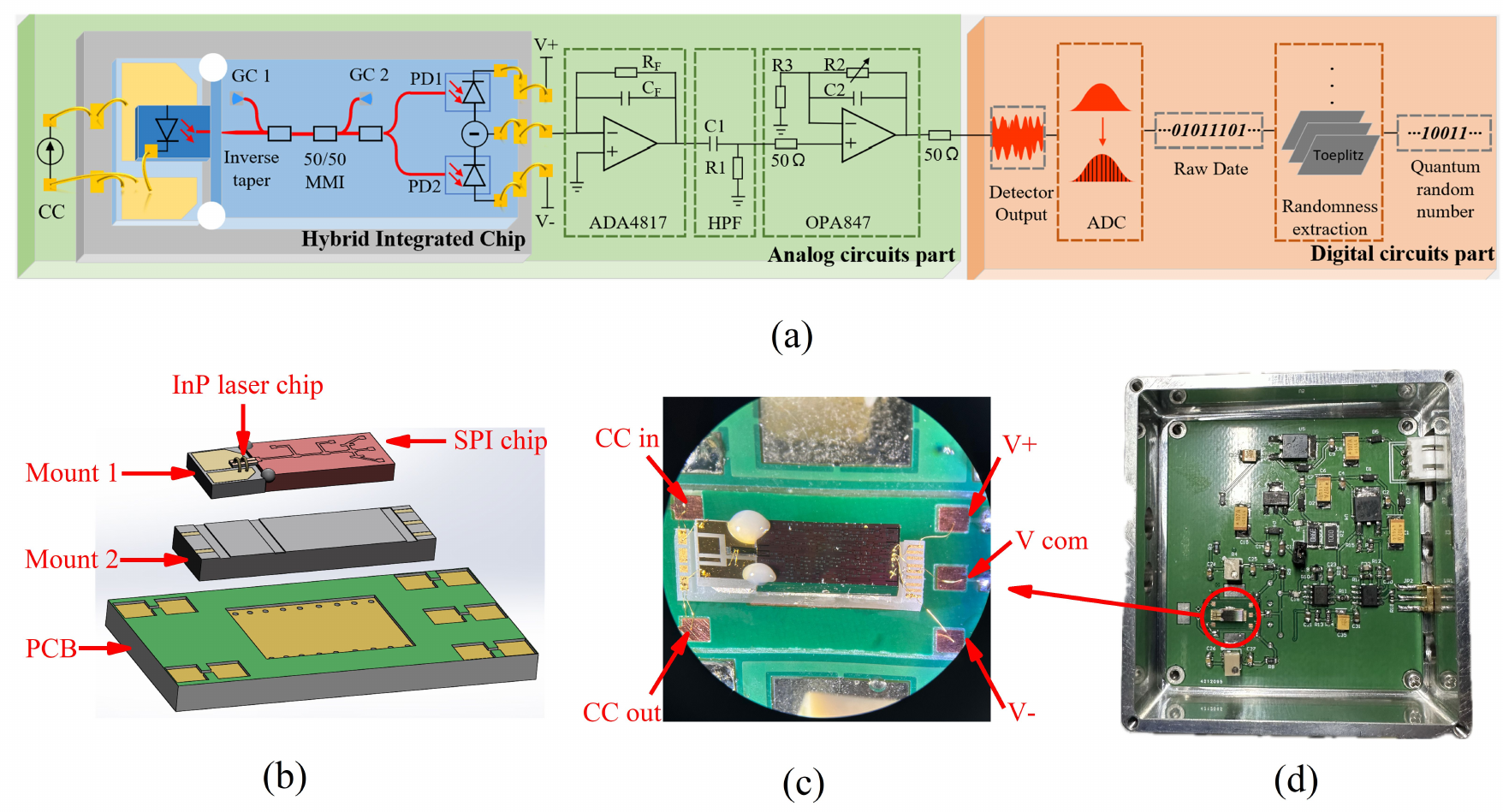}
\caption{  QRNG scheme and images: (a) Scheme, (b) hybrid chip structure, (c) microphotograph of the hybrid chip, and (d) analog circuit. GC: grating coupler, CC: constant current, PD: photodiode, MMI: multimode interferometer, HPF: high-pass filter, PCB: printed circuit board, ADC: analog-to-digital converter, SPI: silicon photonics integrated.}
\end{figure}

In the hybrid chip, one side of the InP LD chip is the anode, and the other side is the cathode. The two electrodes are connected to the CC source by the pads of the two mounts. There are two gold-plated pads on mount 1  (Fig. 1(b)). The anode side of the InP LD chip is mounted on one pad of  mount 1 using the flip-chip method, and the cathode side is connected to the other pad by wire bonding . The InP chip’s size  is 1×0.25 ×0.12 $\text{mm}^3$, with a 40  $\upmu$m gap between it and the edge of mount 1 in the length  or  laser-emitting direction to facilitate the alignment and packaging. Particularly, during packaging, the UV glue can seep into the gap between mount 1 and the silicon photonics chip to make them stick tight, which are then  placed on mount 2. Silver glue is overlaid on mount 2’s surface to fasten the conglutination of mount 1 and the silicon photonics chip. The silver glue should have good heat conduction ability. The packaged hybrid chip is shown in Fig. 1(c). There are also gold-plated welding pads on mount 2. They are used to connect the pads of mount 1 with those of the  printed circuit board (PCB) of the analog circuits part. Fig. 1(d) shows an image of the entire analog circuit.

The CC source in the analog circuit is used to drive the InP LD chip to emit a 1550-nm laser beam with linewidth 0.5 MHz from the hybrid  chip’s edge. Thus, a light beam can be precisely guided into the silicon photonics chip via  edge coupling. The emitted beam polarization is parallel to the silicon photonics chip’s plane and is transformed into a transverse electric mode beam in the waveguide. The insertion loss is  3.1 dB. After packing, the insertion loss increases to 5 dB. To test and package the silicon photonics chip,  two additional grating couplers (GC1,and GC2) are connected to the transmission waveguide. The remaining beam is then split by a 1×2 multimode interferometer (MMI) coupler. The outputs of the two beams are inserted into two GeSi photodiodes. To reduce the system complexity, the 1×2 MMI coupler is designed to achieve a high degree of balance, no Mach-Zehnder interferometer (MZI) structures are used to balance the two output ports of 1×2 MMI\cite{jia2023silicon}.

The silicon photonics chip and analog circuit constitute a typical balance homodyne detector (BHD). In the BHD, only the local oscillator (LO) beam is inserted, the signal beam can be seen as a vacuum beam. The BHD’s output is noise signals, which comprise of quantum and classical noises. To improve the BHD balance degree or common mode rejection ratio (CMRR) , an optimized 1×2 MMI coupler is used. The subtracted photoelectron currents of the two GeSi photodiodes were amplified by a low-noise TIA ADA4817. A feedback resistance with high a value of  $510 \ \text{k} \Omega$ is used. To reduce the feedback circuit’s parasitic capacitance , the PCB under the feedback resistance is hollowed out. The reason for this is described in Section 3. The low-noise voltage amplifier OPA847 is used to magnify the output voltage noise 10–100 times to match the ADC  scale and generate more random numbers.

In the digital circuit, an ADC with 200-kHz samples per second is used to acquire the output voltage noise, and an FPGA is used to extract the random number with a Toeplitz matrix in real time. This process is described in detail in section 4. 

Owing to the low power laser beam, large transimpedance gain, and reasonable second stage selection, no high power LD driver, additional temperature controller,or balance controller is required,  the total consumption power of the entropy source is 80 mW.

\section{QRNG NOISES}
In this section, various QRNG noises  are analyzed and measured. The method for optimizing the QCNR at 1MHz is presented based on a BHD noise model. In this model, all noise densities are transformed into noise current densities, which can be conveniently analyzed. The BHD noises using  commercial InGaAs photodiodes are also analyzed for comparison. The experimental results are consistent with the theoretical analysis. The details are as follows.
\subsection{BHD noise analysis}

\begin{figure}[ht]
\centering\includegraphics[width=12cm]{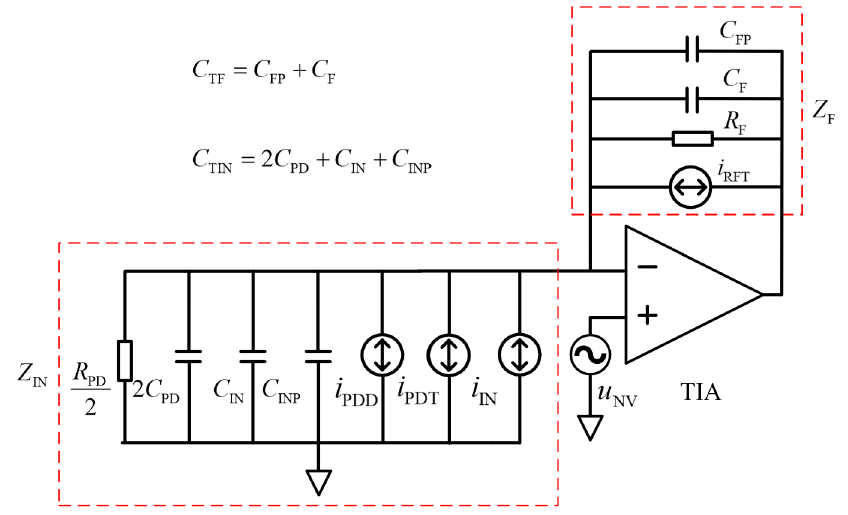}
\caption{ BHD noise model. ${C_{{\rm{TF}}}}$: total capacitance in the feedback loop; ${C_{{\rm{TIN}}}}$: total capacitance in the input port.}
\end{figure}

Figure 2 shows the noise model of the TIA in BHD\cite{wang2017simulation,masalov2017noise}. The measured shunt resistance of the GeSi photodiode is $R_{\text{PD}}=2.47\ \text{M}\Omega$, and the equivalent capacitance of one photodiode is $C_{\text{PD}}=40\ \text{fF}$. The corresponding parameters of the InGaAs photodiode are $10^{11}\ \Omega$ and 0.8 pF. When two photodiodes are in series, as shown in the QRNG scheme in Fig.1, the equivalent shunt resistance in the alternating current (AC) circuit is $R_{\text{PD}}/2$ and the shunt capacitance is $2C_{\text{PD}}$. The input capacitance of the TIA is $C_{\text{IN}}=1.4\ \text{pF}$, and $C_{\text{INP}}$ is the input circuit’s parasitic capacitance.

The thermal noise current density $i_\text{PDT}$ of the GeSi photodiode due to the shunt resistance $R_{\text{PD}}/2$ is given by
\begin{equation}
{i_{{\rm{PDT}}}} = \sqrt {{{4kT} \mathord{\left/
 {\vphantom {{4kT} {\left( {{{{R_{{\rm{PD}}}}} \mathord{\left/
 {\vphantom {{{R_{{\rm{PD}}}}} 2}} \right.
 \kern-\nulldelimiterspace} 2}} \right)}}} \right.
 \kern-\nulldelimiterspace} {\left( {{{{R_{{\rm{PD}}}}} \mathord{\left/
 {\vphantom {{{R_{{\rm{PD}}}}} 2}} \right.
 \kern-\nulldelimiterspace} 2}} \right)}}}  = 1.155 \times {10^{ - 13}}{\rm{ }}{{\rm{A}} \mathord{\left/
 {\vphantom {{\rm{A}} {\sqrt {{\rm{Hz}}} }}} \right.
 \kern-\nulldelimiterspace} {\sqrt {{\rm{Hz}}} }},
\end{equation}
where $k$ is the Boltzmann constant, and $T=295 \ \text{K}$ is the laboratory temperature. ${i_{{\rm{PDT}}}}$ is drawn as the orange solid line in Fig. 3(a). The thermal noise of resistance is white noise, and is invariant with  frequency. For comparison, the thermal noise current density of the InGaAs photodiode, drawn as the orange line in Fig. 3(b), is  $5.739\times10^{-16}\text{A}/\sqrt{\text{Hz}}$ at $T=295 \ \text{K}$. 

\begin{figure}[ht]
\centering\includegraphics[width=12cm]{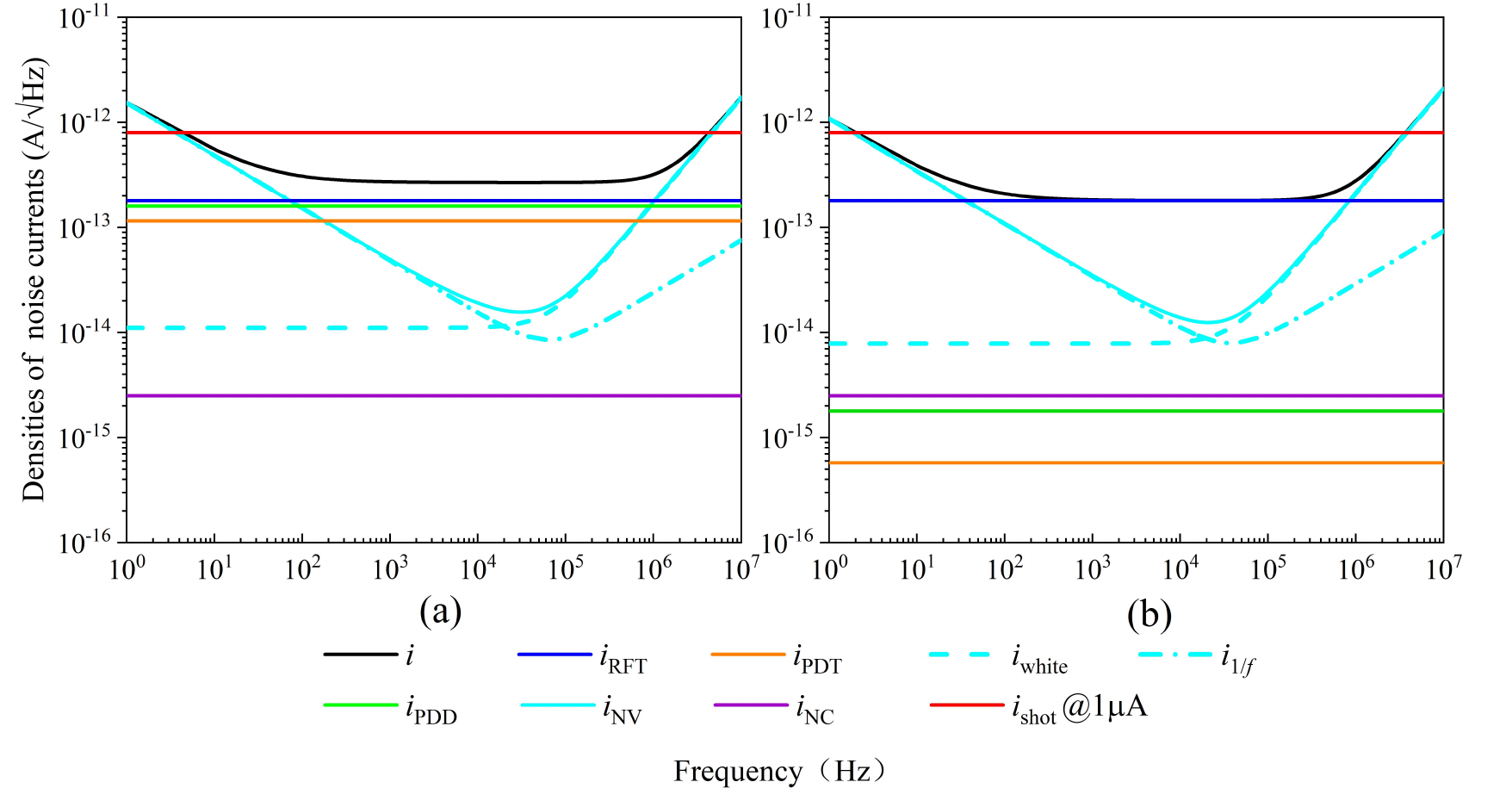}
\caption{ Noise current densities  versus frequency in BHD based on (a) GeSi, and (b) InGaAs photodiodes.}
\end{figure}

The noise current density  $i_{\text{PDD}}$ due to the GeSi photodiode dark current $I_{\text{PDD}}$ can be calculated as follows:
\begin{equation}
i_{\text{PDD}}=\sqrt{2e(2\cdot I_\text{PDD})}.
\end{equation}

The dark current of the GeSi photodiode ${I_{{\rm{PDD}}}}$, drawn as the green solid line in Fig. 3(a), is 4$ \times {10^{ - 8}}$ A when the bias voltage is 2 V. The noise current density ${i_{{\rm{PDD}}}}$, drawn as the green line in Fig. 3(a), is  $1.790 \times {10^{ - 13}}{\rm{ }}{{\rm{A}} \mathord{\left/
 {\vphantom {{\rm{A}} {\sqrt {{\rm{Hz}}} }}} \right.
 \kern-\nulldelimiterspace} {\sqrt {{\rm{Hz}}} }}{\rm{ }}$  and is invariant with frequency. Correspondingly, the dark current of the InGaAs photodiode is  , and the noise current density, drawn as the green solid line in Fig. 3(b), is $1.790 \times {10^{ - 15}}{\rm{ }}{{\rm{A}} \mathord{\left/
 {\vphantom {{\rm{A}} {\sqrt {{\rm{Hz}}} }}} \right.
 \kern-\nulldelimiterspace} {\sqrt {{\rm{Hz}}} }}{\rm{ }}$.

The feedback resistance is ${R_{\rm{F}}} = 510\ {\rm{ k}}\Omega $ , and  ${C_{\rm{F}}}$ is the feedback circuit’s capacitance.  ${C_{{\rm{FP}}}}$ is the feedback circuit’s parasitic capacitance. When the temperature is $T = 295\ {\rm{ K}}$ , the thermal noise current density  ${i_{{\rm{RFT}}}}$ due to feedback resistance, drawn as the blue solid lines in Fig. 3, is 
\begin{equation}
i_{\text{RFT}}=\sqrt{4kT/R_\text{F}}=1.797\times10^{-13}\text{A}/\sqrt{\text{Hz}}.
\end{equation}
It has no relationship with photodiodes and is invariant with frequency. 
The noise current density of the TIA, drawn as the purple solid lines in Fig. 3, is ${i_{{\rm{NC}}}} = 2.5 \times {\rm{1}}{{\rm{0}}^{ - 15}}{{\rm{A}} \mathord{\left/
 {\vphantom {{\rm{A}} {\sqrt {{\rm{Hz}}} }}} \right.
 \kern-\nulldelimiterspace} {\sqrt {{\rm{Hz}}} }}$.

The density of the noise current of the TIA is $i_\text{NC}=2.5\times10^{-15}\text{A}/\sqrt{\text{Hz}}$. It was drawn as the purple solid lines in Fig. 3(a) and Fig. 3(b).

The noise voltage density of the TIA  ${u_{{\rm{NV}}}}$ comprises two parts. The first is the white noise voltage, and its density is
\begin{equation}
u_\text{white}=4  \ \text{nV}/\sqrt{\text{Hz}}.
\end{equation}
The second is  $1/f$ noise voltage, and its density is 
\begin{equation}
u_{1/f}=553.25 \ \text{nV}/\sqrt{f}.
\end{equation}
The total noise voltage density of the TIA is  
\begin{equation}
u_\text{NV}=\sqrt{u_\text{white}^2+u_{1/f}^2}.
\end{equation}
Its contribution to the noise current density  ${i_{{\rm{NV}}}}$ is
\begin{equation}
i_\text{NV}=(1/Z_\text{IN}+Z_\text{F})\cdot u_\text{NV},
\end{equation}
where ${Z_{{\rm{IN}}}}$ denotes the total input impedance of the amplifier, given by
\begin{equation}
Z_\text{IN}=\frac{1}{2/R_\text{PD}+2{\pi}if(2C_\text{PD}+C_\text{INP}+C_\text{IN})},
\end{equation}
and  ${Z_{\rm{F}}}$ is the total feedback impedance of the amplifier, given by
\begin{equation}
Z_\text{F}=\frac{1}{1/R_\text{F}+2{\pi}if(C_\text{F}+C_\text{FP})}.
\end{equation}

The detailed derivation of Eqs. 7-9 is presented in Appendix A. ${i_{{\rm{NV}}}}$ is drawn as cyan solid lines in Fig. 3. The cyan dashed lines represent the white noise current density ${i_{{\rm{white}}}}$, and the cyan dashed-dotted lines represent the  $1/f$ noise current density ${i_{1/f}}$. The two densities vary with frequency. The increase in  ${i_{{\rm{white}}}}$ with frequency is mainly due to the decrease in impedance ${1 \mathord{\left/
 {\vphantom {1 {\left( {{1 \mathord{\left/
 {\vphantom {1 {{Z_{{\rm{IN}}}} + 1/{Z_{\rm{F}}}}}} \right.
 \kern-\nulldelimiterspace} {{Z_{{\rm{IN}}}} + 1/{Z_{\rm{F}}}}}} \right)}}} \right.
 \kern-\nulldelimiterspace} {\left( {{1 \mathord{\left/
 {\vphantom {1 {{Z_{{\rm{IN}}}} + 1/{Z_{\rm{F}}}}}} \right.
 \kern-\nulldelimiterspace} {{Z_{{\rm{IN}}}} + 1/{Z_{\rm{F}}}}}} \right)}}$. 

We now compare the five density types. For the InGaAs photodiode, the thermal noise current density of the photodiode shunt resistance ${i_{{\rm{PDT}}}}$  contributes the least noise and can be neglected. The shunt resistance of the GeSi photodiode is much higher than that of the InGaAs photodiode, and its density ${i_{{\rm{PDT}}}}$  cannot be neglected.The noise current density ${i_{{\rm{PDT}}}}$ of dark current has a similar situation. The noise current density of the TIA ${i_{{\rm{NC}}}}$  can be neglected for both BHD. 

The total noise current density, drawn as the black solid lines in Fig. 3, can be calculated as follows: 
\begin{equation}
i=\sqrt{i_\text{PDT}^2+i_\text{PDD}^2+i_\text{NC}^2+i_\text{RFT}^2+i_\text{NV}^2}.
\end{equation}
Because of the density ${i_{{\rm{NV}}}}$ , the black solid lines can be divided into three stages. In the first stage, the frequency ranges from direct current (DC) to 1 kHz, and the total density decreases with increasing frequency. In the second state, the frequency ranges from 1 kHz to 1 MHz, and the total density is flat. In the third stage, the frequency is greater than 1 MHz, and the total density increases with frequency. In Fig. 3(a), the black lines are slightly higher than the blue lines in the second state (flatness region) because of the contribution of the dark current and shunt resistance of the photodiode. The sum of the above five noise types is usually noted as electronics or classical noise. It is different from the quantum or shot noise generated by the photoelectron current ${I_{{\rm{PD}}}}$ . Red solid lines in Fig. 3 represent the shot noise current density ${i_{shot}}$ generated by the 1 $\upmu$A photoelectron current of the series photodiodes. It can be calculated as follows:
\begin{equation}
i_{shot}=\sqrt{2e(2\cdot I_\text{PD})}=8.006\times10^{-13}.
\end{equation}
It is approximately 9 dB higher than the electronic noise in the flatness region, or the QCNR is 9 dB. 

In the GeSi photodiode-based BHD, to achieve a maximum QCNR and sufficient amplification and bandwidth, a reasonable feedback resistance of 510\ k$\Omega$ is selected. In this case,  $i_\text{RFT}$ is nearly equal to $i_\text{PDD}$ , and  $i_\text{NV}$ is negligible in the flatness region. In this case, a 3-dB bandwidth of approximately 1.2 MHz can be achieved.

In the above analysis, we transform various noises into the noise current densities to compare their amplitudes. The output noise voltage density can be obtained by multiplying the corresponding noise current density by the magnitude of the TIA gain $\left| {G(f)} \right|$. The total noise voltage density can then be obtained as follows:
\begin{equation}
u=i \cdot |G(f)|,
\end{equation}
where  $G(f) $ denotes the TIA gain considering the gain bandwidth product (GBW) of the TIA. It is expressed as follows:
\begin{equation}
G(f)\approx -\frac{1}{\frac{1}{Z_{\text{F}}}+\frac{if}{\text{GBW}}(\frac{1}{Z_\text{F}}+\frac{1}{Z_\text{IN}})}=-\frac{R_\text{F}}{1+\frac{if}{\text{GBW}}+\frac{ifR_\text{F}}{\text{GBW}}\frac{2}{R_\text{PD}}+2\pi ifC_\text{TF}R_\text{F}-\frac{2\pi f^2R_\text{F}}{\text{GBW}}C_\text{TIN}}.
\end{equation}

The GBW is $410\ {\rm{ MHz}}$. A detailed derivation of $G(f) $  can be seen in Appendix A. When the input current is a DC, the frequency $f$  is zero and the gain is $- {R_{\rm{F}}}$ . 

\begin{figure}[ht]
\centering\includegraphics[width=12cm]{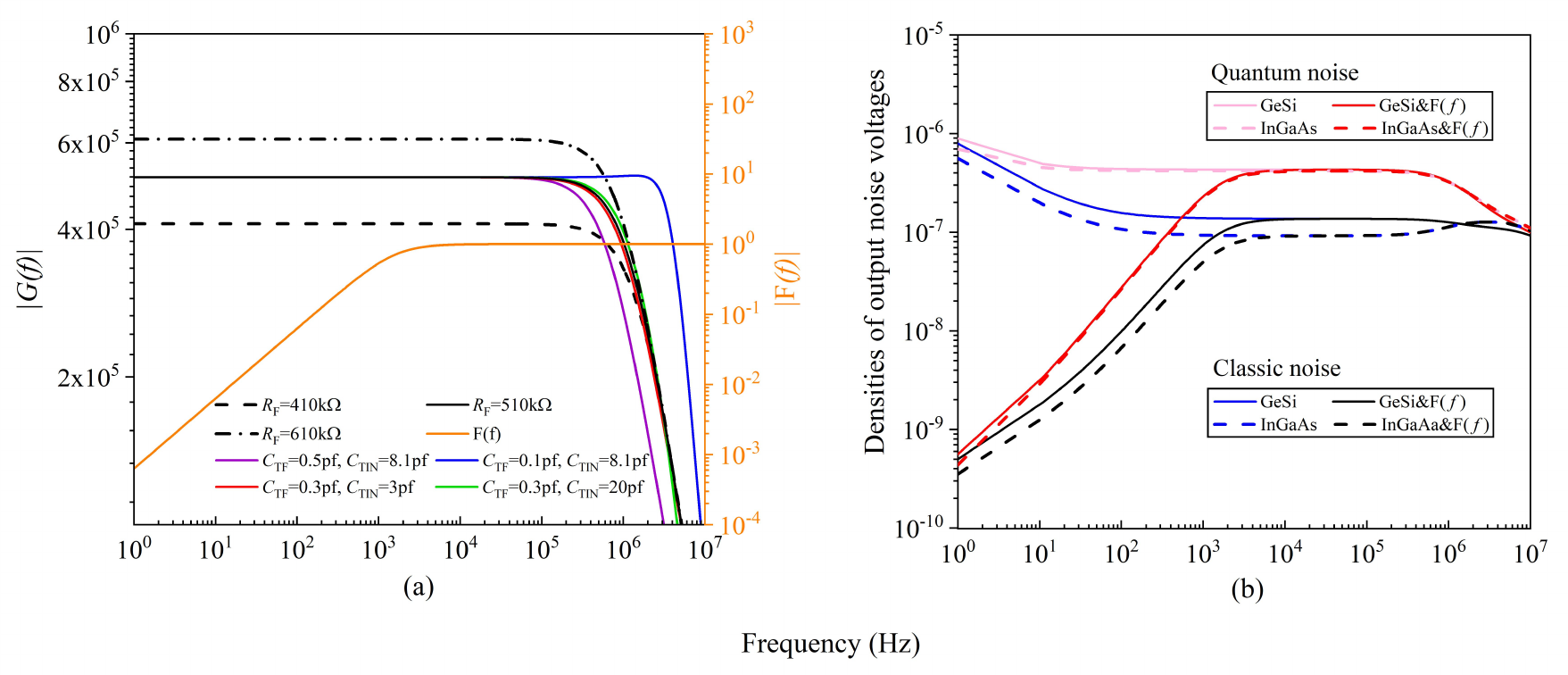}
\caption{ Transimpedance gain mode and output noise voltage densities versus frequency: (a) transimpedance gain mode  $\left| {G(f)} \right|$ and (b) output noise voltage densities.}
\end{figure}

The curves of  $\left| {G(f)} \right|$ are shown in Fig. 4(a). The value of  ${R_{\rm{F}}}$ determines the gain amplitude of the flatness region. When the value of ${R_{\rm{F}}}$  is larger, the gain value in the flatness region is larger. In this case,  ${i_{{\rm{NV}}}}$ is smaller and the bandwidth of  $\left| {G(f)} \right|$ is smaller. In the experiment, we should choose the value of ${R_{\rm{F}}}$  to ensure that ${i_{{\rm{RFT}}}}$  and  ${i_{{\rm{NV}}}}$ are as small as possible and the 3-dB bandwidth is larger than 1 MHz. The bandwidth is mainly determined by ${R_{\rm{F}}}$ ,${C_{{\rm{TF}}}}$ , and ${C_{{\rm{TIN}}}}$ . We set  ${C_{{\rm{TF}}}}$ to 0.1, 0.3, and 0.5 pf, and the curves are the solid blue, balck, and purple lines in Fig. 4(a), respectively. When  ${C_{{\rm{TF}}}}$ is smaller, the bandwidth is broader. We also set ${C_{{\rm{TIN}}}}$  to 3, 8.1, and 20 pF, and the curves are the red, black, and green lines in Fig. 4(b), respectively. The variance in bandwidth is not obvious. Thus, we do not use ${C_{\rm{F}}}$  and hollow the PCB below  ${R_{\rm{F}}}$ to minimize ${C_{{\rm{FP}}}}$. Due to the parasitic capacitance  ${C_{{\rm{FP}}}}$ and ${C_{{\rm{INP}}}}$  can’t be measured in experiment, they are determined by comparising the calculated noise power and the measured noise power introduced in section 3.2. The capacitance  ${C_{{\rm{TF}}}}$ and  ${C_{{\rm{TIN}}}}$ of GeSi based BHD are determined as 0.3 pf and 8.1 pf. Then the parasitic capacitance  ${C_{{\rm{FP}}}} = {C_{{\rm{TF}}}}$ is 0.3 pF, and ${C_{{\rm{INP}}}} = {C_{{\rm{TIN}}}} - 2 \times {C_{{\rm{PD}}}} - {C_{{\rm{IN}}}}$  is 6.62 pF.  

The orange solid line in Fig. 4(a) represents the transmission function  ${\rm{F}}(f)$ of the HPF. The 3-dB bandwidth is 1.6 kHz. The HPF is used to filter the low-frequency noise in which there is mainly $1/f$  noise and the DC signal.

The classical output noise voltage density of the GeSi photodiode-based BHD is drawn as the black solid line in Fig. 4(b), and the blue solid line represents the density before the HPF. The red solid curve represents the output density of the sum of quantum and classical noise voltages when the photoelectron current is 1 $\upmu$A, and the pink solid line represents the density before the HPF. The dashed lines represent the corresponding results for the InGaAs photodiode-based BHD. In the flatness region, the red solid line nearly overlaps with the corresponding red dashed line and the black solid line is slightly higher than the corresponding black dashed line. 
\subsection{BHD noise power measurement}
The noise power measurement results, which are consistent with the theoretically calculated results, are presented in detail in this section. When the output impedance of the TIA is 50  $\Omega$ and the input impedance of the radio frequency (RF) spectrum analyzer is $R = 50{\rm{ }}\Omega $ , the noise voltage density $u$  can be transformed into the noise power density, $s$ , as follows: 
\begin{equation}
s \ {\rm{ }}{{{\rm{dBm}}} \mathord{\left/
 {\vphantom {{{\rm{dBm}}} {{\rm{Hz}}}}} \right.
 \kern-\nulldelimiterspace} {{\rm{Hz}}}} = 10 \ {\log _{10}}\left( {\frac{{{{\left( {{u \mathord{\left/
 {\vphantom {u 2}} \right.
 \kern-\nulldelimiterspace} 2}} \right)}^2}}}{R} \cdot \frac{1}{{1 \times {{10}^{ - 3}}{\rm{W}}}}} \right).
\end{equation}

The unit of  $u$  is  ${{\rm{V}} \mathord{\left/
 {\vphantom {{\rm{V}} {\sqrt {{\rm{Hz}}} }}} \right.
 \kern-\nulldelimiterspace} {\sqrt {{\rm{Hz}}} }}$ and the unit of $R$ is $\Omega $. The noise power $S$  with the resolution bandwidth $\Delta f$  is 
\begin{equation}
S \ =s\cdot \Delta f \ \text{dBm}.
\end{equation}

\begin{figure}[ht]
\centering\includegraphics[width=12cm]{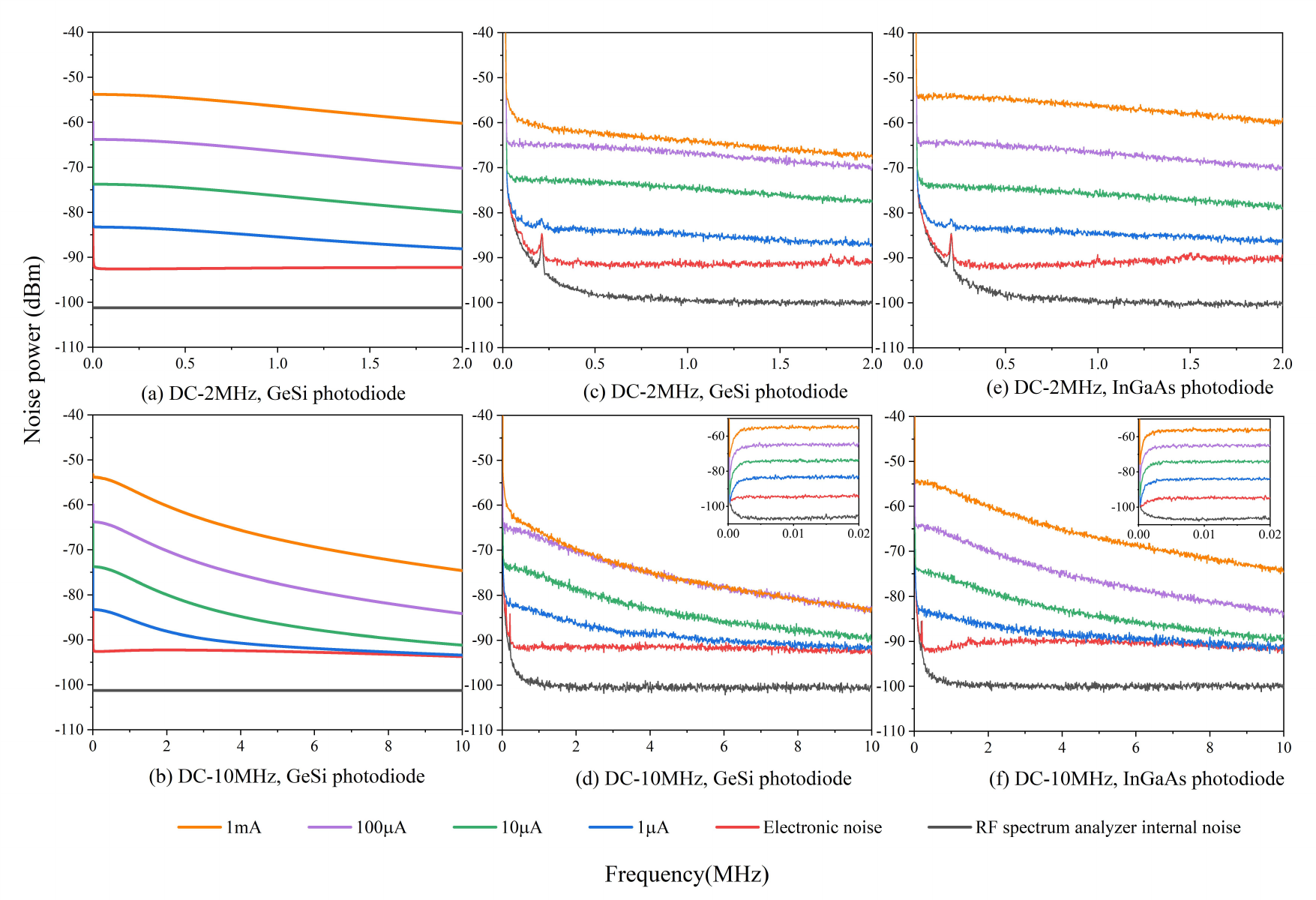}
\caption{ Calculated and measured noise power of BHD based on two types of photodiodes.  Calculated results with GeSi photodiode: (a) DC to 2 MHz and (b) DC to 10 MHz. Measurement results with GeSi photodiode: (c) DC to 2 MHz and (d) DC to 10 MHz. Measurement results with InGaAs photodiode: (c) DC to 2 MHz and (d) DC to 10 MHz. }
\end{figure}

Fig. 5(a) shows the calculated noise power at different photoelectron currents in the frequency range from DC to 2 MHz, with a resolution bandwidth of 10 kHz. The black spectrum curve represents the internal noise power of the RF spectrum analyzer without considering the low-frequency noise. The blue spectrum curve represents the electronics or classical noise power of the GeSi photodiode-based BHD. The green spectrum curve represents the noise power when the photoelectron current is 1 $\upmu$A. A 9-dB signal-to-noise ratio can be achieved. The 3-dB bandwidth is 1.2 MHz. The green, purple, and orange spectrum curves represent the noise power when the photoelectron currents are 10 $\upmu$A, 100 $\upmu$A, and 1 mA, respectively. The interval of the noise power is 10 dB. The spectrum scale was extended from 2 to 10 MHz (Fig. 5(b)).

Figures 5(c) and (d) show the measurement results obtained using the BHD based on the GeSi photodiode of the hybrid chip. The values of the curves are consistent with those of the curves in Figs. 5(a) and (b), except for the effect due to the low-frequency noise of the RF spectrum. For comparison, the experimental results of the InGaAs photodiode-based BHD are also shown in Figs. 5(e) and (f). The noise power based on the two types of photodiodes is nearly the same. The main difference is that the maximum QCNR of the GeSi photodiode-based BHD is lower, attributable to the non-balancing structure in the hybrid chip. When the photoelectron current is large, the difference in photoelectron currents causes saturation of the BHD. Thus, the noise power of 1 mA in Figs. 5(c) and (d) is abnormal. Inset figures in Figs. 5 (d) and (f) present low-frequency noise. To decrease the effect of the low-frequency noise of the RF spectrum analyzer, the TIA output noise was magnified 10 times by the voltage amplifier OPA847 and the resolution bandwidth was changed to 100 Hz. From the spectrum, the detectors are shot noise limited in the entire magnified range of the BHD. 

\subsection{CMRR}
\begin{figure}[ht]
\centering\includegraphics[width=12cm]{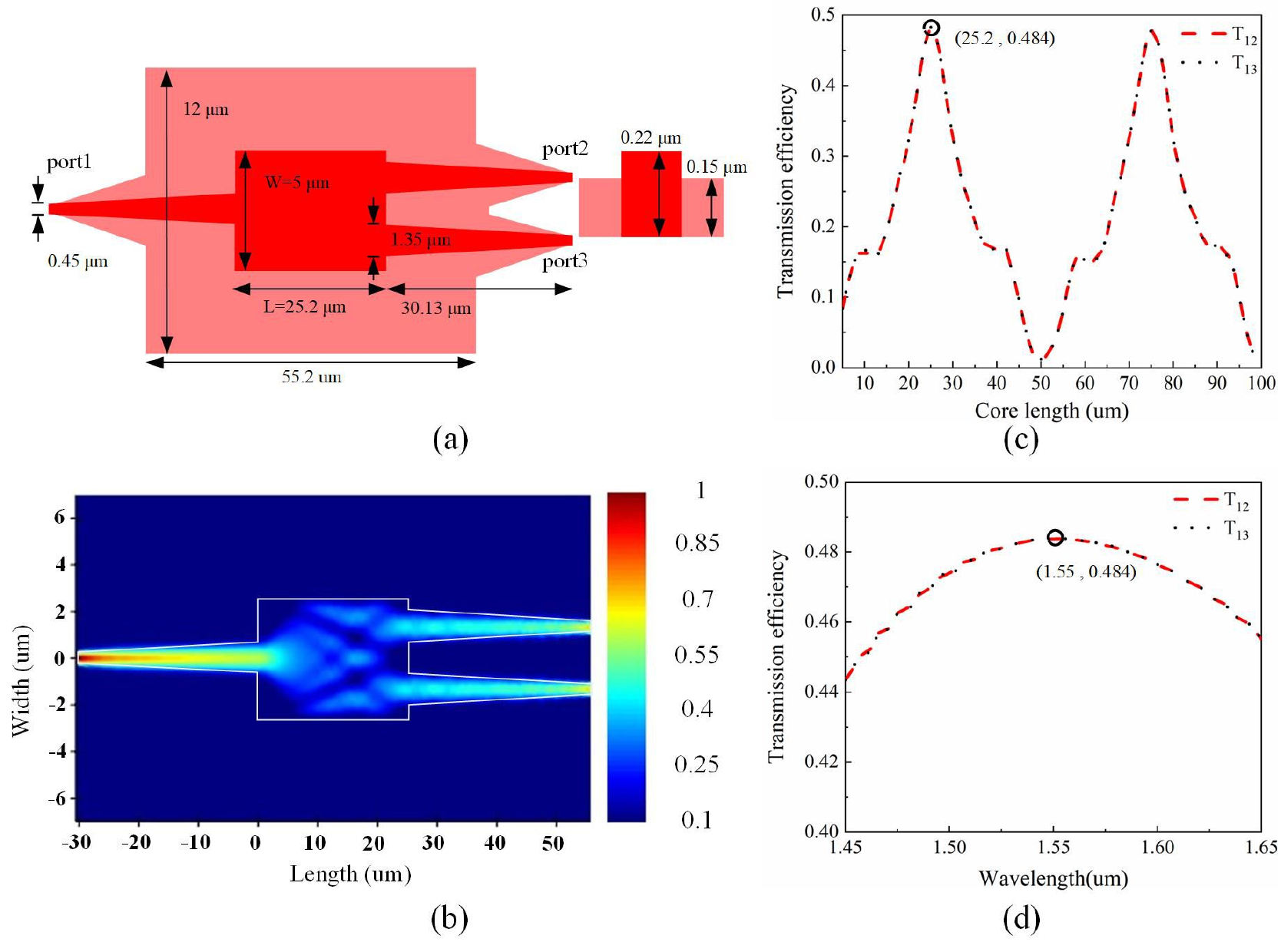}
\caption{ Optimized 1×2 MMI structure: (a) size and (b) transmission efficiency versus core length. (c) Simulated power distribution at a 1550-nm wavelength and (d) transmission efficiency versus wavelength.}
\end{figure}
A high-CMRR BHD can cancel the intensity noise of the laser\cite{jin2015balanced,wang2023accurate}. Usually, to achieve a high CMRR, silicon photonics chip-based BHDs use the MZI or p-i-n phase modulator structure to balance the two output ports of a 50/50 MMI coupler\cite{jia2023silicon}. The additional balance structure requires extra balance control, which will perplex the BHD structure and QRNG, and more cost and power will be required. To achieve a high degree of balance in situations without a balance structure, we designed and optimized a symmetrical 1×2 MMI structure (Fig. 6) to achieve as high a balance degree as possible. The balance of the two output beams is more insensitive to the size and wavelength than that of the 2×2 MMI structure. Besides, the size of 1×2 MMI is smaller and the loss is lower. Overall, it is more suitable for use in QRNG than 2×2 MMI\cite{siew2021review}.

The size of the optimized 1×2 MMI structure is shown in Fig. 6(a). The core length is $L = 25.2 \ {\rm{ \upmu m}}$  and core width $W = 5 \ {\rm{ \upmu m}}$.  Fig. 6(b) depicts the transmission efficiency versus the core length. The transmission efficiency from port 1 to ports 2 and 3 is  ${T_{12}}$ and ${T_{13}}$, respectively. The variance is periodic, and we select the first peak point with the highest transmission efficiency. The top view of the simulated power distribution at a 1550-nm wavelength is shown in Fig. 6(c). The wavelength is also optimized according to the 1550-nm laser beam (Fig. 6(d)). Even if the size may have an error in fabrication or the laser beam may change, the balance will be unaffected.

\begin{figure}[ht]
\centering\includegraphics[width=12cm]{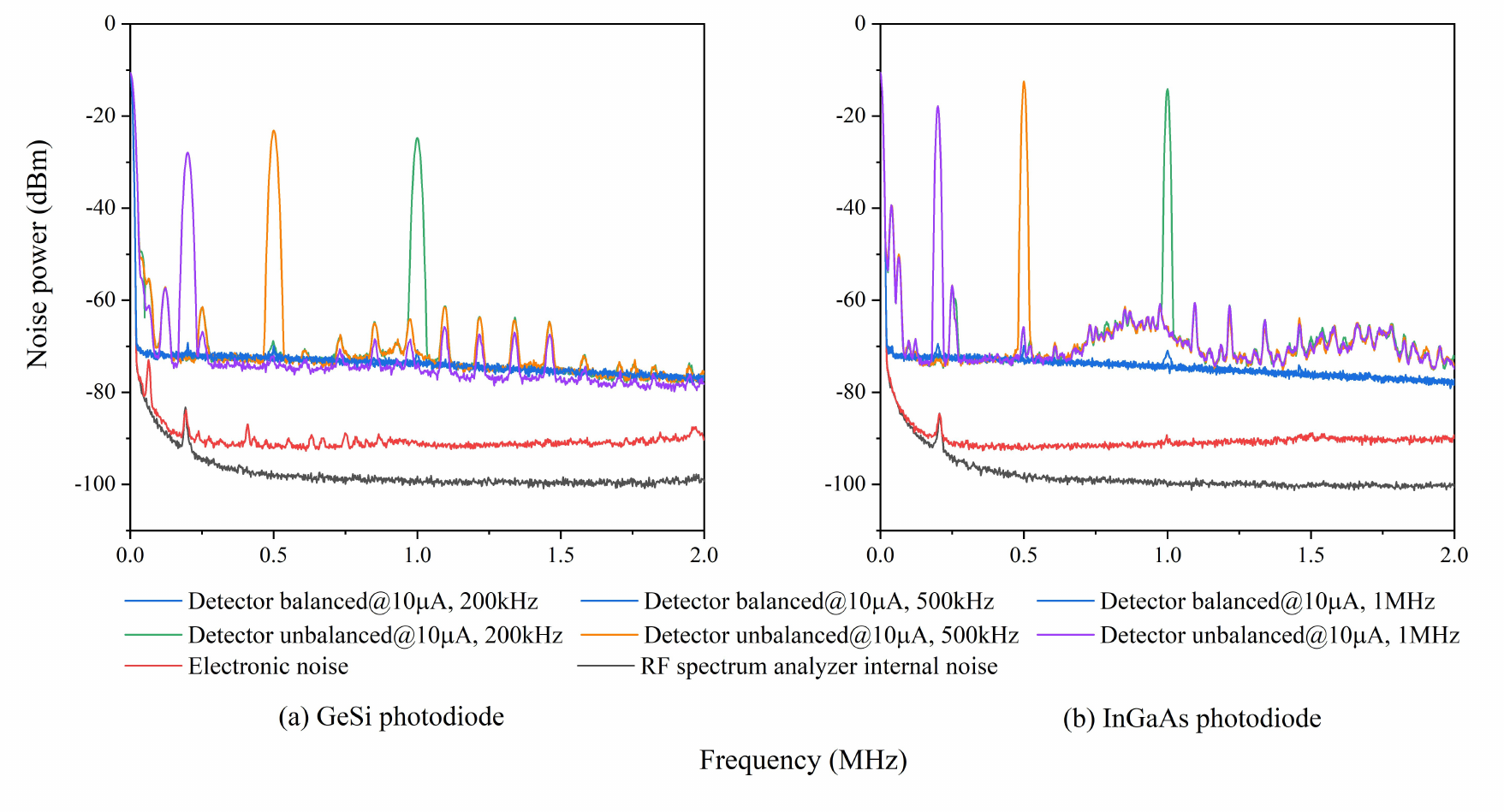}
\caption{ CMRR measurement results of BHD based on (a) GeSi photodiode and (b) InGaAs photodiode.}
\end{figure}

From the measurement results in Fig. 7(a), a high CMRR greater than 40 dB can be achieved, the photoelectron current is 10 $\upmu$A, and the measured frequencies are 200 kHz, 500 kHz, and 1 MHz. Figure 7(b) shows the measurement results of the CMRR of the InGaAs photodiode-based BHD for comparison. Because the intensities of the two output ports of the BHD can be tunable, the CMRR is higher and the measurement result is greater than 50 dB.

In CMRR measurements, it is inconvenient to directly modulate the amplitude of the laser beam emitted by the InP laser chip. We guide the laser beam emitted by the other InP laser chip into an amplitude modulator and then guide the modulated beam into the silicon photonics chip using GC1. In this process, the shell of the shield metal box is open, with a small disturbance in the electronic noise (Fig. 7(a)).

\section{GENERATION OF QUANTUM RANDOM NUMBER}
From the above experimental results, silicon photonics integrated and shot noise-limited BHD can be achieved. It is the foundation for generating quantum random numbers. We now employ the average conditional minimum entropy ${\bar H_{\min }}$ \cite{haw2015maximization}, which gives the eavesdropper maximum power, including an infinite ADC range ${R_e} \to \infty $  and an infinitely small bin ${\delta _e} \to 0$. It is assumed that an adversary can only listen but has no control over the classical noise.  ${\bar H_{\min }}$ is expressed as follows: 
\begin{equation}
\begin{array}{c}
{{\bar H}_{\min }}\left( {{M_{{\rm{dis}}}}|E} \right) = \mathop {\lim }\limits_{{\delta _e} \to 0} {{\bar H}_{\min }}\left( {{M_{{\rm{dis}}}}|{E_{{\rm{dis}}}}} \right)\\
 =  - {\log _2}\left[ {\int_{ - \infty }^{ + \infty } {{P_E}\left( e \right)} \mathop {\max }\limits_{{m_i} \in {M_{{\rm{dis}}}}} {P_{{M_{{\rm{dis}}}}|E}}\left( {{m_i}|e} \right)de} \right],
\end{array}
\end{equation}
 where $m$ denotes the measured total output noise data, and  $e$ denotes the classical noise data. ${M_{{\rm{dis}}}}$  and  ${E_{{\rm{dis}}}}$ are their probability distributions. For more details about Eq. (16), refer to a previous study\cite{haw2015maximization} .

\begin{figure}[ht]
\centering\includegraphics[width=9cm]{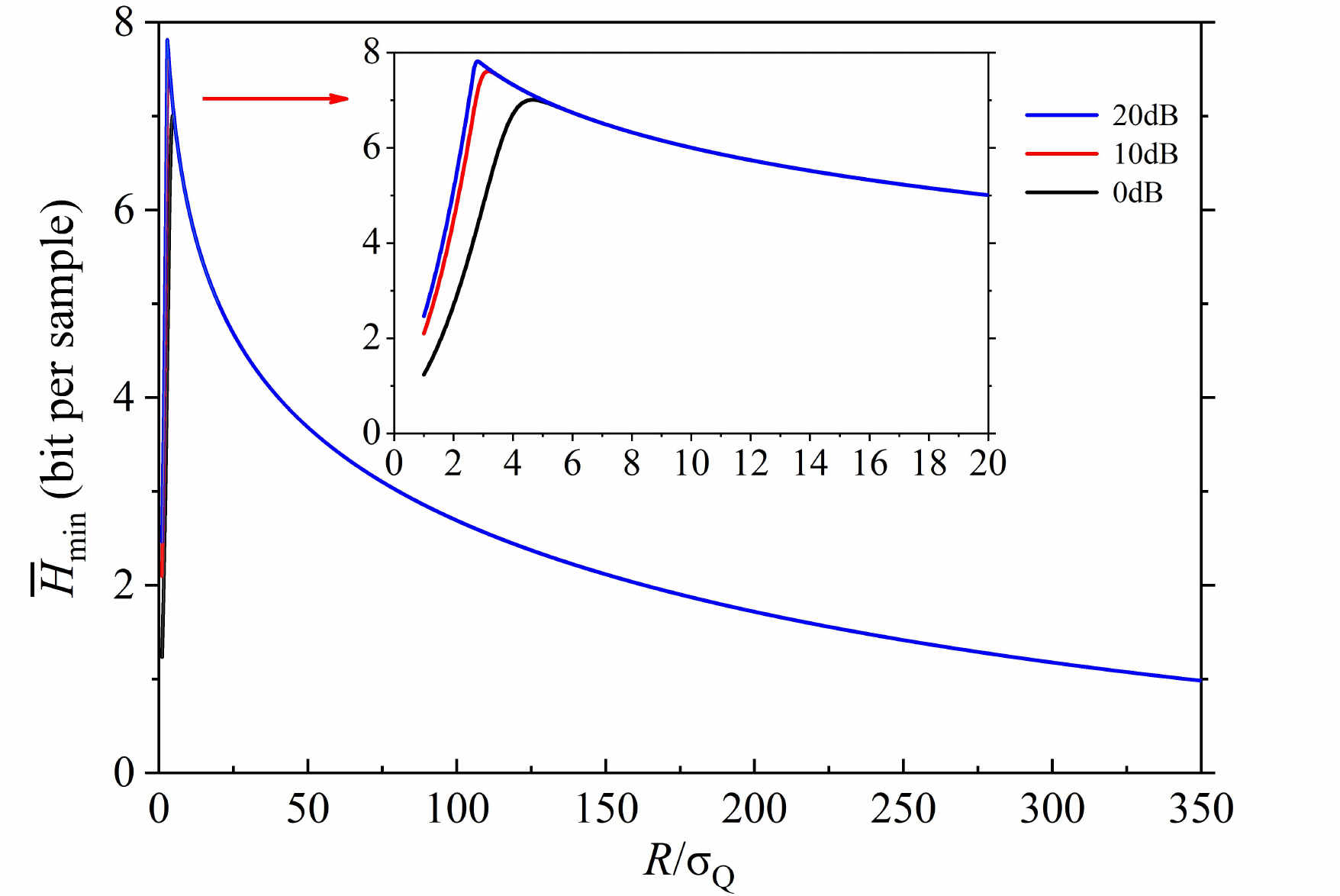}
\caption{Average conditional minimum entropy versus the ratio ${R \mathord{\left/
 {\vphantom {R {{\sigma _{\rm{Q}}}}}} \right.
 \kern-\nulldelimiterspace} {{\sigma _{\rm{Q}}}}}$ }
\end{figure}

In the experiment, a common ADC range  $ \pm R =  \pm {\rm{5 V}}$ is used, and the resolution is 8 bits. With the numerical method, ${\bar H_{\min }}$ versus the ratio ${R \mathord{\left/
 {\vphantom {R {{\sigma _{\rm{Q}}}}}} \right.
 \kern-\nulldelimiterspace} {{\sigma _{\rm{Q}}}}}$ was drawn at different QCNRs (Fig. 8), where ${\sigma _{\rm{Q}}}$  denotes the SD of the quantum noise.  

\begin{table}[!htbp]
\caption{SDs at different magnifications and photoelectron currents}
\label{table:1}
\centering
\begin{tabular}{ >{\centering\arraybackslash}m{2.2cm} 
>{\centering\arraybackslash}m{2.2cm} 
>{\centering\arraybackslash}m{2.2cm} 
>{\centering\arraybackslash}m{2.2cm}
>{\centering\arraybackslash}m{2.2cm}
}
 \hline Magnification & ${\sigma _{\rm{E}}}\ \text{(V)}$ & ${\sigma _{\rm{Q}}}@1 \ \upmu \text{A (V)}$ & ${\sigma _{\rm{Q}}}@10 \  \upmu \text{A (V)}$ & ${\sigma _{\rm{Q}}}@100 \ \upmu \text{A (V)}$ \\ 
 \hline 10 & 0.0058 & 0.0153 & 0.0474 & 0.1537\\ 
 20 & 0.0112 & 0.0267 & 0.0933 &  0.2923\\ 
 50 & 0.0282 & 0.0746 & 0.2068 &  0.6607\\ 
 100 & 0.0566  & 0.1386 & 0.4402 & 1.3764\\ 
\hline
 \end{tabular}
\end{table}

 From the inset figure, when the QCNR increases, the optimized ratio ${R \mathord{\left/
 {\vphantom {R {{\sigma _{\rm{Q}}}}}} \right.
 \kern-\nulldelimiterspace} {{\sigma _{\rm{Q}}}}}$  decreases. If we choose the peak point to calculate ${\bar H_{\min }}$, a large  ${\sigma _{\rm{Q}}}$ and high magnification are required. For example, when the QCNR is 20 dB, the optimized ratio is ${R \mathord{\left/
 {\vphantom {R {{\sigma _{\rm{Q}}}}}} \right.
 \kern-\nulldelimiterspace} {{\sigma _{\rm{Q}}}}} =2.3$. Then, an SD of ${\sigma _{\rm{Q}}} =2.17$ is required. From Table 1, when the laser beam power is 100 $\upmu$A and the magnification is 100 times, ${\sigma _{\rm{Q}}} =1.376$, which is still smaller than the required 2.17. Under realistic conditions, we should choose a suitable QCNR ratio considering the laser beam power, magnification, required random number generation rate, and power together. In the experiment, we decrease  ${R \mathord{\left/
 {\vphantom {R {{\sigma _{\rm{Q}}}}}} \right.
 \kern-\nulldelimiterspace} {{\sigma _{\rm{Q}}}}}$ to 20. In this case, although ${R \mathord{\left/
 {\vphantom {R {{\sigma _{\rm{Q}}}}}} \right.
 \kern-\nulldelimiterspace} {{\sigma _{\rm{Q}}}}}$ increases to 20 and  ${\sigma _{\rm{Q}}} =0.25$ is required. The required second magnification decreases to approximately 20 when the photoelectron current is 100 $\upmu$A. When the photoelectron current is 10 $\upmu$A, a magnification between 50 and 100 is required. We can select the parameters flexibly according to the actual application in which the MHz quantum random number generation rate is sufficient.

Based on the selected parameters above, the measured noise with  ${\sigma _{\rm{M}}} = \sqrt {\sigma _{\rm{E}}^2 + \sigma _{\rm{Q}}^2}  = 0.27\ {\rm{ V}}$ and the electronics noise with ${\sigma _{\rm{E}}} = 0.028\ {\rm{ V}}$  are drawn with red and black points, respectively, in Fig. 9. Figure 9 (b) shows a histogram of the acquired data with a Gaussian distribution. The sample rate was 200 kHz. The calculated ${\bar H_{\min }}$ using Eq. (16) is 5.117 bps. Data processing is based on an FPGA card in real time, which uses the Toeplitz matrix to extract the random number according to ${\bar H_{\min }}$ \cite{lu2021quantum}. Finally, a quantum random number generation rate of 1.02 megabits per second can be achieved. 
 
\begin{figure}[ht]
\centering\includegraphics[width=13cm]{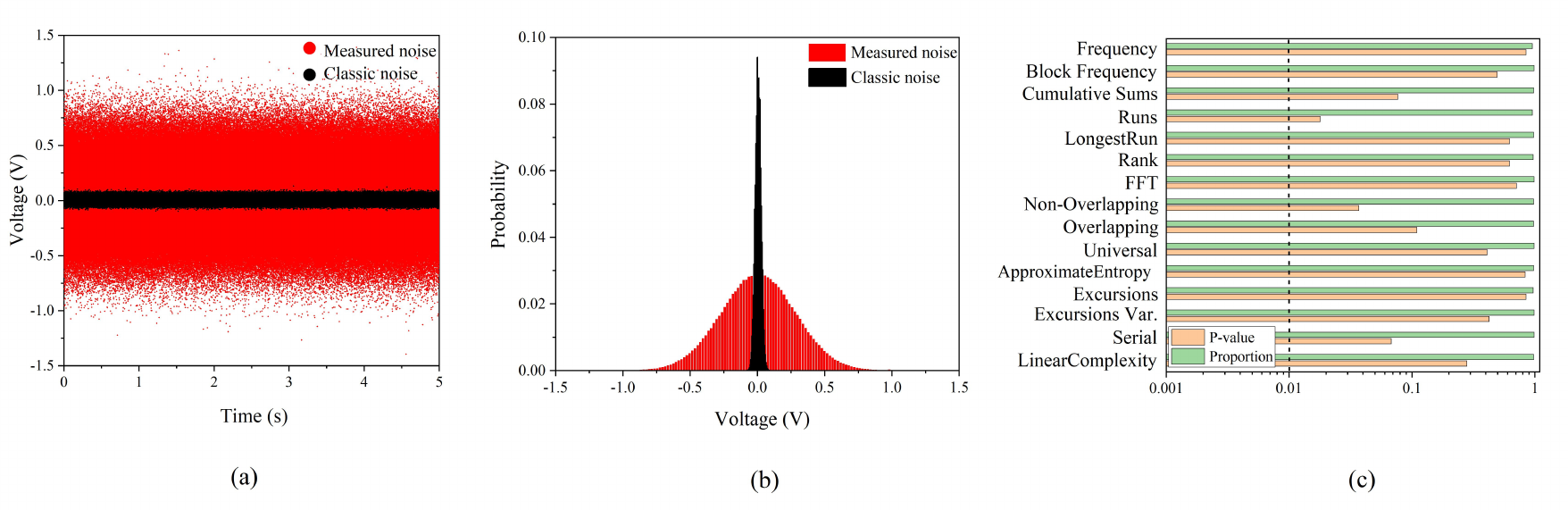}
\caption{(a) Time traces of obtained quantum and classical noise data, (b) histogram of obtained quantum and classical noise data, and (c) randomness test result.}
\end{figure}

The NIST SP 800-22 suite was employed to test randomness. The total amount of extracted random numbers used for the test is 1 Gigabit. The results in Fig. 9 (c) demonstrate that the random numbers successfully passed the test, indicating that the generated random sequence has good statistical characteristics. 

 \section{CONCLUSION AND OUTLOOK}

 In conclusion, we have reported a compact and low power consumption vacuum noise QRNG based on a hybrid chip comprising a commercial 1550-nm InP LD chip and a silicon photonics integrated chip fabricated using industry-standard active flow SOI technology. The packaged hybrid chip’s size is 8.8×2.6×1 $\text{mm}^3$. A method for maximizing the QCNR at MHz is presented. The calculated noise power is consistent with the noise measurement results. A QCNR of approximately 9 dB can be achieved when the photoelectron current is 1 $\upmu$A. Because a high QCNR can be achieved when a low-power laser beam is injected, no high-value CC or temperature controller is used. Although the balancing structure is deleted, a CMRR greater than 40 dB can still be achieved with an optimized 1×2 MMI coupler. Finally, the total power consumption of the entropy source is 80 mW. A tunable and high gain voltage amplifier is used to flexibly tune the SD of the output noise. The required SD according to the input voltage scale of the ADC is also analyzed. The SD of the output noise voltage is adapted to the ADC in the digital circuit. A tunable and high gain voltage amplifier is used to flexibly tune the SD of the output noise. The required SD according to the input voltage scale of the ADC is also analyzed. QRNG has the potential for use in a scenario of moderate MHz random number generation speed, with low power, small volume, and low cost prioritized.  
 
This study explores a technique for implementing a compact and low-cost QRNG module for practical applications. In the future, a higher generation speed is expected to be obtained based on the hybrid chip by increasing the TIA bandwidth. Further, we will integrate the optical and electrical components on one chip based on standard CMOS manufacturing processes. Consequently, a more compact and low-cost QRNG can be realized.

\section*{Funding}
This project was supported by the Provincial Natural Science Foundation of Shanxi, China (Grant No.202103021224010), Shanxi Provincial Foundation for Returned Scholars, China (Grant No.2022-016), Aeronautical Science Foundation of China (Grant No.20200020115001), National Natural Science Foundation of China (Grant Nos.62175138, 62205188, 11904219), the Program of State Key Laboratory of Quantum Optics and Quantum Optics Devices (Grant No. KF202006), the “1331 Project” for Key Subject Construction of Shanxi Province, China, and the Innovation
Program for Quantum Science and Technology (2021ZD0300703).

\section*{Appendix A}
\begin{figure}[ht]
\centering\includegraphics[width=12cm]{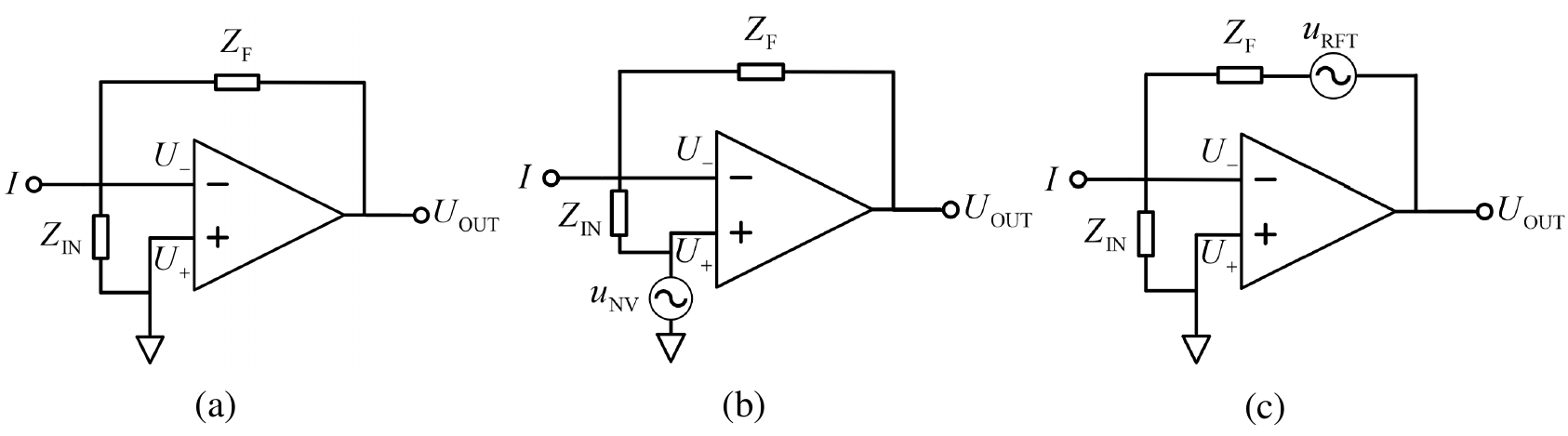}
\caption{Typical TIA circuit based on an operation amplifier with different noise : (a) without noise, (b) with noise voltage of TIA ${u_{{\rm{NV}}}}$, and (c) with noise voltage of feedback resistance ${u_{{\rm{RFT}}}}$.}
\end{figure}

Figure 10(a) shows the typical TIA based on the operation amplifier without noise;  $I$ denotes the input current,  ${Z_{{\rm{IN}}}}$ denotes the total input impedance, and  ${Z_{\rm{F}}}$ denotes the feedback impedance. The GBW is  ${A_0}{f_0}$ , which is the product of the open loop voltage gain ${A_0}$  and measurement frequency ${f_0}$ . Usually, the open loop gain of the amplifier can be expressed as follows: 
\begin{equation}
A(f) = \frac{{{A_0}}}{{1 + if/{f_0}}} = \frac{{{U_{{\rm{OUT}}}}}}{{{U_ + } - {U_ - }}}.
\end{equation}
According to Kirchhoff’s current law, 

\begin{equation}
I = \frac{{{U_ - }}}{{{Z_{{\rm{IN}}}}}} + \frac{{{U_ - } - {U_{{\rm{OUT}}}}}}{{{Z_{\rm{F}}}}}.
\end{equation}
Using Eqs. (17) and (18), the TIA gain can be calculated as follows:

\begin{equation}
\begin{array}{c}
G(f) = \frac{{{U_{out}}}}{I} =  - \frac{{A(f){U_ - }}}{{\frac{{{U_ - }}}{{{Z_{{\rm{IN}}}}}} + \frac{{{U_ - } - {U_{out}}}}{{{Z_{\rm{F}}}}}}} =  - \frac{1}{{\frac{1}{{{Z_{\rm{F}}}}} + \frac{1}{{A\left( f \right)}}(\frac{1}{{{Z_{\rm{F}}}}} + \frac{1}{{{Z_{{\rm{IN}}}}}})}}\\
 =  - \frac{1}{{\frac{1}{{{Z_{\rm{F}}}}} + \frac{1}{{{A_0}}}(\frac{1}{{{Z_{\rm{F}}}}} + \frac{1}{{{Z_{{\rm{IN}}}}}}) + \frac{{if}}{{{A_0}{f_0}}}(\frac{1}{{{Z_{\rm{F}}}}} + \frac{1}{{{Z_{{\rm{IN}}}}}})}}.
\end{array}
\end{equation}

In the above derivation, the voltage ${U_ + }$  is set to zero. Because ${A_0}$  is very large, the term

\begin{equation}
\frac{1}{{{A_0}}}(\frac{1}{{{Z_{\rm{F}}}}} + \frac{1}{{{Z_{{\rm{IN}}}}}}) \ll \frac{1}{{{Z_{\rm{F}}}}}
\end{equation}
can be neglected. The gain can then be simplified as follows:

\begin{equation}
G(f) =  - \frac{1}{{\frac{1}{{{Z_{\rm{F}}}}} + \frac{{if}}{{{\rm{GBW}}}}(\frac{1}{{{Z_{\rm{F}}}}} + \frac{1}{{{Z_{{\rm{IN}}}}}})}}.
\end{equation}
When the input current of the circuit is zero, Eq. (18) can be rewritten as follows:

\begin{equation}
{U_{{\rm{OUT}}}} = \frac{{{Z_{{\rm{IN}}}} + {Z_{\rm{F}}}}}{{{Z_{{\rm{IN}}}}}}{U_ - }.
\end{equation}
The noise voltage source can be placed at the positive input port of the TIA (Fig. 10 (b)). In this case, ${U_ + } = {u_{{\rm{NV}}}}$ . Using Eqs. (17) and (22), we obtain 

\begin{equation}
{U_{{\rm{OUT}}}} = \frac{1}{{{1 \mathord{\left/
 {\vphantom {1 {A + {{{Z_{{\rm{IN}}}}} \mathord{\left/
 {\vphantom {{{Z_{{\rm{IN}}}}} {\left( {{Z_{{\rm{IN}}}} + {Z_{\rm{F}}}} \right)}}} \right.
 \kern-\nulldelimiterspace} {\left( {{Z_{{\rm{IN}}}} + {Z_{\rm{F}}}} \right)}}}}} \right.
 \kern-\nulldelimiterspace} {A + {{{Z_{{\rm{IN}}}}} \mathord{\left/
 {\vphantom {{{Z_{{\rm{IN}}}}} {\left( {{Z_{{\rm{IN}}}} + {Z_{\rm{F}}}} \right)}}} \right.
 \kern-\nulldelimiterspace} {\left( {{Z_{{\rm{IN}}}} + {Z_{\rm{F}}}} \right)}}}}}}{u_{{\rm{NV}}}}.
\end{equation}
Using Eqs. (19), (21), and (22), we obtain 

\begin{equation}
{U_{{\rm{OUT}}}} = \left( {{1 \mathord{\left/
 {\vphantom {1 {{Z_{{\rm{IN}}}} + {1 \mathord{\left/
 {\vphantom {1 {{Z_{\rm{F}}}}}} \right.
 \kern-\nulldelimiterspace} {{Z_{\rm{F}}}}}}}} \right.
 \kern-\nulldelimiterspace} {{Z_{{\rm{IN}}}} + {1 \mathord{\left/
 {\vphantom {1 {{Z_{\rm{F}}}}}} \right.
 \kern-\nulldelimiterspace} {{Z_{\rm{F}}}}}}}} \right){u_{{\rm{NV}}}}G\left( f \right) = {i_{{\rm{NV}}}}G\left( f \right).
\end{equation}
The thermal noise voltage of the feedback resistance is shown in Fig. 10(c), where the noise voltage density of the feedback resistance is ${v_{{\rm{RFT}}}} = \sqrt {4kT{R_{\rm{F}}}} $ . 
According to Kirchhoff’s current law, 

\begin{equation}
\frac{{{U_ - }}}{{{Z_{{\rm{IN}}}}}} = \left( {{U_{out}} - {U_ - }} \right) \cdot 2\pi if{C_{\rm{F}}} + \frac{{{U_{out}} - {U_ - } - {u_{{\rm{RFT}}}}}}{{{R_{\rm{F}}}}}.
\end{equation}
Using Eq. (17) and setting ${U_ + } = 0$ , we obtain 

\begin{equation}
{U_{{\rm{OUT}}}} = \sqrt {{{4kT} \mathord{\left/
 {\vphantom {{4kT} {{R_{\rm{F}}}}}} \right.
 \kern-\nulldelimiterspace} {{R_{\rm{F}}}}}} G\left( f \right) = {i_{{\rm{RFT}}}}G\left( f \right).
\end{equation}




\begin{thebibliography}{10}
	\newcommand{\enquote}[1]{``#1''}
	
	\bibitem{RevModPhys.89.015004}
	M.~Herrero-Collantes and J.~C. Garcia-Escartin, \enquote{Quantum random number
		generators,} {\protect\JournalTitle{Rev. Mod. Phys.}} \textbf{89}(1), 015004
	(2017).
	
	\bibitem{2016Quantum}
	X.~Ma, X.~Yuan, Z.~Cao, \emph{et~al.}, \enquote{Quantum random number
		generation,} {\protect\JournalTitle{npj Quantum Information}} \textbf{2}(1), 16021
	(2016).
	
	\bibitem{2006Random}
	H.~Bauke and S.~Mertens, \enquote{Random numbers for large-scale distributed
		monte carlo simulations,} {\protect\JournalTitle{Physical Review E
			Statistical Nonlinear \& Soft Matter Physics}} \textbf{75}, 066701 (2006).
	
	\bibitem{Lim2015Quantum}
	Lim, Charles, Ci, \emph{et~al.}, \enquote{Quantum random number generation for
		1.25-GHz quantum key distribution systems,} {\protect\JournalTitle{Journal of
			Lightwave Technology A Joint IEEE/OSA Publication}} \textbf{33}(13), 2855-2859 (2015).
	
	\bibitem{liu2023experimental}
	S.~Liu, Z.~Lu, P.~Wang, \emph{et~al.}, \enquote{Experimental demonstration of
		multiparty quantum secret sharing and conference key agreement,}
	{\protect\JournalTitle{npj Quantum Information}} \textbf{9}(1), 92 (2023).
	
	\bibitem{2023High}
	Y.~Tian, Y.~Zhang, S.~Liu, \emph{et~al.}, \enquote{High-performance
		long-distance discrete-modulation continuous-variable quantum key
		distribution,} {\protect\JournalTitle{Optics Letters}} \textbf{48}(11), 2953-2956 (2023).
	
	\bibitem{2000A}
	T.~Jennewein, U.~Achleitner, G.~Weihs, \emph{et~al.}, \enquote{A fast and
		compact quantum random number generator,} {\protect\JournalTitle{Review of
			Scientific Instruments}} \textbf{71}(4), 1675--1680 (2000).
	
	\bibitem{2007Quantum}
	M.~Stipcevic and B.~M. Rogina, \enquote{Quantum random number generator based
		on photonic emission in semiconductors,} {\protect\JournalTitle{Review of
			Scientific Instruments}} \textbf{78}(4), 045104 (2007).
	
	\bibitem{2009Photon}
	M.~A. Wayne, E.~R. Jeffrey, G.~M. Akselrod, and P.~G. Kwiat, \enquote{Photon
		arrival time quantum random number generation,}
	{\protect\JournalTitle{Journal of Modern Optics}} \textbf{56}(4), 516--522
	(2009).
	
	\bibitem{wahl2011ultrafast}
	M.~Wahl, M.~Leifgen, M.~Berlin, \emph{et~al.}, \enquote{An ultrafast quantum
		random number generator with provably bounded output bias based on photon
		arrival time measurements,} {\protect\JournalTitle{Applied Physics Letters}}
	\textbf{98}(17), 145-266 (2011).
	
	\bibitem{furst2010high}
	H.~F{\"u}rst, H.~Weier, S.~Nauerth, \emph{et~al.}, \enquote{High speed optical
		quantum random number generation,} {\protect\JournalTitle{Optics Express}}
	\textbf{18}(12), 13029--13037 (2010).
	
	\bibitem{ren2011quantum}
	M.~Ren, E.~Wu, Y.~Liang, \emph{et~al.}, \enquote{Quantum random-number
		generator based on a photon-number-resolving detector,}
	{\protect\JournalTitle{Physical Review A}} \textbf{83}(2), 023820 (2011).
	
	\bibitem{wei2009bias}
	W.~Wei and H.~Guo, \enquote{Bias-free true random-number generator,}
	{\protect\JournalTitle{Optics letters}} \textbf{34}(12), 1876--1878 (2009).
	
	\bibitem{gabriel2010generator}
	C.~Gabriel, C.~Wittmann, D.~Sych, \emph{et~al.}, \enquote{A generator for
		unique quantum random numbers based on vacuum states,}
	{\protect\JournalTitle{Nature Photonics}} \textbf{4}(10), 711--715 (2010).
	
	\bibitem{shen2010practical}
	Y.~Shen, L.~Tian, and H.~Zou, \enquote{Practical quantum random number
		generator based on measuring the shot noise of vacuum states,}
	{\protect\JournalTitle{Physical Review A}} \textbf{81}(6), 063814 (2010).
	
	\bibitem{symul2011real}
	T.~Symul, S.~M. Assad, and P.~K. Lam, \enquote{Real time demonstration of high
		bitrate quantum random number generation with coherent laser light,}
	{\protect\JournalTitle{Applied Physics Letters}} \textbf{98}(23), 145 (2011).
	
	\bibitem{guo2010truly}
	H.~Guo, W.~Tang, Y.~Liu, and W.~Wei, \enquote{Truly random number generation
		based on measurement of phase noise of a laser,}
	{\protect\JournalTitle{Physical Review E}} \textbf{81}, 051137 (2010).
	
	\bibitem{qi2010high}
	B.~Qi, Y.-M. Chi, H.-K. Lo, and L.~Qian, \enquote{High-speed quantum random
		number generation by measuring phase noise of a single-mode laser,}
	{\protect\JournalTitle{Optics letters}} \textbf{35}(3), 312--314 (2010).
	
	\bibitem{jofre2011true}
	M.~Jofre, M.~Curty, F.~Steinlechner, \emph{et~al.}, \enquote{True random
		numbers from amplified quantum vacuum,} {\protect\JournalTitle{Optics
			Express}} \textbf{19}(21), 20665--20672 (2011).
	
	\bibitem{williams2010fast}
	C.~R. Williams, J.~C. Salevan, X.~Li, \emph{et~al.}, \enquote{Fast physical
		random number generator using amplified spontaneous emission,}
	{\protect\JournalTitle{Optics Express}} \textbf{18}(23), 23584--23597 (2010).
	
	\bibitem{bustard2011quantum}
	P.~J. Bustard, D.~Moffatt, R.~Lausten, \emph{et~al.}, \enquote{Quantum random
		bit generation using stimulated raman scattering,}
	{\protect\JournalTitle{Optics Express}} \textbf{19}(25), 25173--25180 (2011).
	
	\bibitem{marandi2011twin}
	A.~Marandi, N.~C. Leindecker, K.~L. Vodopyanov, and R.~L. Byer, \enquote{Twin
		degenerate opo for quantum random bit generation,} in \emph{Nonlinear Optics:
		Materials, Fundamentals and Applications,}  (Optica Publishing Group, 2011),
	p. NME4.
	
	\bibitem{bruynsteen2023100}
	C.~Bruynsteen, T.~Gehring, C.~Lupo, \emph{et~al.}, \enquote{100-Gbit/s
		integrated quantum random number generator based on vacuum fluctuations,}
	{\protect\JournalTitle{PRX Quantum}} \textbf{4}, 010330 (2023).
	
	\bibitem{raffaelli2018homodyne}
	F.~Raffaelli, G.~Ferranti, D.~H. Mahler, \emph{et~al.}, \enquote{A homodyne
		detector integrated onto a photonic chip for measuring quantum states and
		generating random numbers,} {\protect\JournalTitle{Quantum Science and
			Technology}} \textbf{3}, 025003 (2018).
	
	\bibitem{tasker2021silicon}
	J.~F. Tasker, J.~Frazer, G.~Ferranti, \emph{et~al.}, \enquote{Silicon photonics
		interfaced with integrated electronics for 9 Gz measurement of squeezed
		light,} {\protect\JournalTitle{Nature Photonics}} \textbf{15}(1), 11--15 (2021).
	
	\bibitem{bruynsteen2021integrated}
	C.~Bruynsteen, M.~Vanhoecke, J.~Bauwelinck, and X.~Yin, \enquote{Integrated
		balanced homodyne photonic--electronic detector for beyond 20 GHz
		shot-noise-limited measurements,} {\protect\JournalTitle{Optica}} \textbf{8}(9),
	1146--1152 (2021).
	
	\bibitem{bai202118}
	B.~Bai, J.~Huang, G.-R. Qiao, \emph{et~al.}, \enquote{18.8 Gbps real-time
		quantum random number generator with a photonic integrated chip,}
	{\protect\JournalTitle{Applied Physics Letters}} \textbf{118}, 264001 (2021).
	
	\bibitem{abellan2016quantum}
	C.~Abellan, W.~Amaya, D.~Domenech, \emph{et~al.}, \enquote{Quantum entropy
		source on an inp photonic integrated circuit for random number generation,}
	{\protect\JournalTitle{Optica}} \textbf{3}(9), 989--994 (2016).
	
	\bibitem{siew2021review}
	S.~Y. Siew, B.~Li, F.~Gao, \emph{et~al.}, \enquote{Review of silicon photonics
		technology and platform development,} {\protect\JournalTitle{Journal of
			Lightwave Technology}} \textbf{39}(13), 4374--4389 (2021).
	
	\bibitem{kaur2021hybrid}
	P.~Kaur, A.~Boes, G.~Ren, \emph{et~al.}, \enquote{Hybrid and heterogeneous
		photonic integration,} {\protect\JournalTitle{APL Photonics}} \textbf{6}, 061102
	(2021).
	
	\bibitem{jia2023silicon}
	Y.~Jia, X.~Wang, X.~Hu, \emph{et~al.}, \enquote{Silicon photonics-integrated
		time-domain balanced homodyne detector in continuous-variable quantum key
		distribution,} {\protect\JournalTitle{New J. Phys}} \textbf{25}, 103030
	(2023).
	
	\bibitem{wang2017simulation}
	S.~Wang, X.~Xiang, C.~Zhou, \emph{et~al.}, \enquote{Simulation of high snr
		photodetector with lc coupling and transimpedance amplifier circuit and its
		verification,} {\protect\JournalTitle{Review of Scientific Instruments}}
	\textbf{88}, 013107 (2023).
	
	\bibitem{masalov2017noise}
	A.~Masalov, A.~Kuzhamuratov, and A.~Lvovsky, \enquote{Noise spectra in balanced
		optical detectors based on transimpedance amplifiers,}
	{\protect\JournalTitle{Review of Scientific Instruments}} \textbf{88}, 113109 (2017).
	
	\bibitem{jin2015balanced}
	X.~Jin, J.~Su, Y.~Zheng, \emph{et~al.}, \enquote{Balanced homodyne detection
		with high common mode rejection ratio based on parameter compensation of two
		arbitrary photodiodes,} {\protect\JournalTitle{Optics Express}} \textbf{23}(18),
	23859--23866 (2015).
	
	\bibitem{wang2023accurate}
	X.-Y. Wang, X.-B. Guo, Y.-X. Jia, \emph{et~al.}, \enquote{Accurate
		shot-noise-limited calibration of a time-domain balanced homodyne detector
		for continuous-variable quantum key distribution,}
	{\protect\JournalTitle{Journal of Lightwave Technology}} \textbf{41}(17), 5518-5528 (2023).
	
	\bibitem{haw2015maximization}
	J.-Y. Haw, S.~Assad, A.~Lance, \emph{et~al.}, \enquote{Maximization of
		extractable randomness in a quantum random-number generator,}
	{\protect\JournalTitle{Physical Review Applied}} \textbf{3}(5), 054004 (2015).
	
	\bibitem{lu2021quantum}
	Z.~Lu, J.~Liu, X.~Wang, \emph{et~al.}, \enquote{Quantum random number generator
		with discarding-boundary-bin measurement and multi-interval sampling,}
	{\protect\JournalTitle{Optics Express}} \textbf{29}(8), 12440--12453 (2021).
	
\end{thebibliography}
\normalem

\end{document}